\documentclass[10pt,aps,twocolumn,prl,nofootinbib,noshowkeys,superscriptaddress,floatfix]{revtex4-1}

\usepackage{amsmath,amsfonts,amsthm,amssymb}
\usepackage[dvips]{graphics,graphicx}
\usepackage{nicefrac}
\usepackage[usenames,dvipsnames]{color}
\definecolor{darkblue}{RGB}{0,0,196}
\definecolor{darkgreen}{RGB}{0,120,0}
\usepackage[colorlinks=true,linktocpage=true,linkcolor=darkblue,citecolor=red,urlcolor=darkblue]{hyperref}
\usepackage{cancel}
\usepackage{multirow}
\usepackage{longtable}
\usepackage{color}
\usepackage[normalem]{ulem}
\usepackage{hyperref}
\usepackage{bigints}
\usepackage{yfonts}
\usepackage{xparse}
\usepackage{physics}
\usepackage{verbatim}
\usepackage{minibox}
\usepackage{comment}
\usepackage{appendix}

\def\spin{\,\textgoth{s:}}

\begin{document}

\preprint{}

\title{Relativistic Spin Magnetohydrodynamics}

\author{Samapan Bhadury}
\email{samapan.bhadury@niser.ac.in}
\affiliation{School of Physical Sciences, National Institute of Science Education and Research, An OCC of Homi Bhabha National Institute, Jatni-752050, India}
\author{Wojciech Florkowski}
\email{wojciech.florkowski@uj.edu.pl}
\affiliation{Institute of Theoretical Physics, Jagiellonian University, ul. St. \L ojasiewicza 11, 30-348 Krakow, Poland}
\author{\!Amaresh Jaiswal}
\email{a.jaiswal@niser.ac.in}
\thanks{Corresponding Author}
\affiliation{School of Physical Sciences, National Institute of Science Education and Research, An OCC of Homi Bhabha National Institute, Jatni-752050, India}
\author{\!Avdhesh Kumar} 
\email{avdheshk@gate.sinica.edu.tw}
\affiliation{School of Physical Sciences, National Institute of Science Education and Research, An OCC of Homi Bhabha National Institute, Jatni-752050, India}
\affiliation{Institute of Physics, Academia Sinica, Taipei, 11529, Taiwan}
\author{Radoslaw Ryblewski} 
\email{radoslaw.ryblewski@ifj.edu.pl}
\affiliation{Institute of Nuclear Physics Polish Academy of Sciences, PL-31-342 Krakow, Poland}
\date{\today}


\begin{abstract}
Starting from kinetic theory description of massive spin-$\nicefrac{1}{2}$ particles in presence of magnetic field, equations for relativistic dissipative non-resistive magnetohydrodynamics are obtained in the small polarization limit. We use a relaxation time approximation for the collision kernel in the relativistic Boltzmann equation and calculate non-equilibrium corrections to the phase-space distribution function of spin-polarizable particles. We demonstrate that our framework naturally leads to emergence of the well known Einstein-de Hass and Barnett effects. We obtain multiple transport coefficients and show, for the first time, that the coupling between spin and magnetic field appear at gradient order in hydrodynamic equations.
\end{abstract}


\maketitle



{\bf \textit{Introduction}}-- In non-central heavy-ion collisions, a fireball with large global angular momentum \cite{Becattini:2007sr} is created which experiences very strong magnetic field \cite{Tuchin:2013ie} at early times due to fast-moving charged spectators. While the angular momentum in the fireball survives for a longer time due to the conservation of total angular momentum, the magnetic field tends to decay rapidly due to fast receding spectators. However, the medium can sustain it for a longer time provided it has large electrical conductivity. Nevertheless, these extreme physical conditions may generate a spin polarization and magnetization of the hot and dense matter, very similar to magneto-mechanical effects of Einstein-de Haas \cite{Einstein} and Barnett \cite{Barnett}. Consequently, it was predicted that signatures of such phenomena in relativistic heavy-ion collision may be found in spin-polarization of observed particles \cite{Liang:2004ph, Liang:2004xn, Voloshin:2004ha, Betz:2007kg, Becattini:2007nd}. Recently, much effort has been devoted to studies of spin-polarization of particles produced in high-energy nuclear collisions, both from the experimental \cite{STAR:2017ckg, STAR:2018gyt,STAR:2019erd, STAR:2020xbm, STAR:2021beb, ALICE:2019aid} and theoretical perspective \cite{Becattini:2013fla, Becattini:2013vja, Li:2017slc, Sun:2017xhx, Wang:2017jpl, Becattini:2018duy, Florkowski:2018ahw, Xia:2018tes, Hattori:2019lfp, Wu:2019eyi, Sheng:2019kmk, Becattini:2019ntv, Yang:2020hri, Deng:2020ygd, Florkowski:2021pkp, Li:2021zwq, Yang:2021fea, Muller:2021hpe, Yi:2021unq, Yi:2021ryh}. 

It has been well established that the fireball formed in high-energy heavy-ion collisions behaves like a fluid \cite{Heinz:2013th}. In order to develop a hydrodynamic framework, which allows for space-time evolution of polarization effects, one needs to consider the conservation of total angular momentum \cite{Florkowski:2018fap, Speranza:2020ilk, Becattini:2020ngo}, along with the usual energy-momentum and net particle current conservation. This has led to the formulation of relativistic spin-hydrodynamics \cite{Florkowski:2017ruc, Bhadury:2020cop, Bhadury:2020puc, Hu:2021pwh, Shi:2020htn, Fu:2020oxj, Speranza:2021bxf, She:2021lhe, Peng:2021ago, Wang:2021ngp, Yi:2021unq, Wang:2021wqq, Florkowski:2019qdp, Singh:2020rht, Singh:2021man, Florkowski:2021wvk, Das:2022azr, Montenegro:2017rbu, Montenegro:2017lvf, Montenegro:2018bcf, Montenegro:2020paq, Gallegos:2021bzp, Hongo:2021ona, Cartwright:2021qpp, Gallegos:2022jow}. Similarly, in order to study the evolution of strong magnetic field produced in high-energy heavy-ion collisions, the theory of relativistic magnetohydrodynamics was also formulated \cite{Roy:2015kma, Pu:2016ayh, Hernandez:2017mch, Florkowski:2018ubm, Denicol:2018rbw, Denicol:2019iyh, Panda:2020zhr, Panda:2021pvq, Karabali:2014vla, Hattori:2016cnt, Hattori:2017qih, Dash:2017rhg, Mohanty:2018eja}. However, these two effects on polarization observables are not entirely separable and therefore a unified framework of “spin-magnetohydrodynamics” needs to be developed \cite{Jaiswal:2020rtc}.

Considering the phenomena of Einstein-de Haas and Barnett effects, it is expected that coupling between spin polarization and magnetization occurs in presence of rotation and/or electromagnetic field. Therefore one needs a formulation of spin-magnetohydrodynamics to incorporate the effect of this coupling. In our previous work on the formulation of dissipative spin-hydrodynamics \cite{Bhadury:2020puc, Bhadury:2020cop}, we showed that the spin tensor acquires contributions from various thermodynamic gradients or forces. Thus, as a consequence, the spin-polarization too receives contribution due to these forces \cite{Liu:2021uhn, Becattini:2021suc}. It is therefore imperative that we generalize our formulation to also include strong magnetic field produced in initial stages of relativistic heavy-ion collisions.

In the present work, we develop such a framework for a single species of massive spin-$\nicefrac{1}{2}$ particles that is electrically charged with finite chemical potential. We obtain the hydrodynamic equations of motion for this system which can exhibit spin-polarization and magnetization and demonstrate that these equations are consistent with macroscopic conservation laws within kinetic theory framework. We consider relativistic Boltzmann equation for such a system in the presence of magnetic field and obtain equations for relativistic dissipative non-resistive magnetohydrodynamics in the limit of small polarization. We obtain multiple transport coefficients and show that dissipative currents contain coupling between spin and magnetic field at first-order in gradients.

We use the convention $g_{\mu\nu} =  \hbox{diag}(+1,-1,-1,-1)$ and $\epsilon^{0123} = -\epsilon_{0123}=1$ for the metric tensor and totally anti-symmetric Levi-Civita symbol, respectively. Throughout the text we use natural units where $c = \hbar = k_B =1$.



{\bf \textit{Equations of motion}}-- In absence of external source of particles, the net particle four-current remains conserved. The particle four-current is given by
\begin{equation}\label{Nmu}
    N^\mu = n u^\mu + n^\mu, 
\end{equation}
where $n$ is the equilibrium net particle number-density, $u^\mu$ is the fluid four-velocity defined in the Landau frame and $n^\mu$ is the the dissipative particle number diffusion. The conservation of particle current implies 
\begin{equation}\label{N^mu_cons}
    \partial_\mu N^\mu = 0
\end{equation}
The stress-energy tensor of the system of fluid and electromagnetic field can be expressed as  \cite{Florkowski:2018ubm}
\begin{align}
    T^{\mu\nu} &= T^{\mu\nu}_{\mathrm{f}} + T^{\mu\nu}_{\mathrm{int}} + T^{\mu\nu}_{\mathrm{em}} . \label{T^mn_decomp}
\end{align}
Here $T^{\mu\nu}_{\mathrm{f}}$ denotes contribution from fluid, $T^{\mu\nu}_{\mathrm{em}}$ denotes contribution from field and $T^{\mu\nu}_{\mathrm{int}}$ denotes interaction between fluid and field.

These three components of the stress-energy tensors can be further expressed as \cite{Israel:1978up, Florkowski:2018ubm},
\begin{align}
    T^{\mu\nu}_{\mathrm{f}} &= \epsilon u^\mu u^\nu - \left(P + \Pi\right) \Delta^{\mu\nu} + \pi^{\mu\nu} \label{T^mn_f}\\
    T^{\mu\nu}_{\mathrm{int}} &= - \Pi^\mu u^\nu - F^{\mu}_{~\,\alpha} M^{\nu\alpha} \label{T^mn_int1}\\
    T^{\mu\nu}_{\mathrm{em}} &= - F^{\mu\alpha} F^{\nu}_{~\,\alpha} + \frac{1}{4} g^{\mu\nu} F^{\alpha\beta} F_{\alpha\beta} \label{T^mn_em}
\end{align}
where, in Eq.~\eqref{T^mn_f} $\epsilon$ and $P$ are the equilibrium energy density and pressure, $\Pi$ and $\pi^{\mu\nu}$ are the bulk and shear viscous pressures, and, $\Delta^{\mu\nu} = g^{\mu\nu} - u^\mu u^\nu$ is the projection operator orthogonal to $u^\mu$. The form of stress-energy tensor in Eq.~\eqref{T^mn_f} is written in Landau frame: $T^{\mu\nu}_{\mathrm f}u_\nu=\epsilon u^\mu$. The additional quantities appearing in Eqs.~\eqref{T^mn_int1} and \eqref{T^mn_em} i.e. $\Pi^\mu$, $F^{\mu\nu}$ and $M^{\mu\nu}$ are an auxiliary four-vector, Maxwell field strength tensor and electromagnetic polarization-magnetization tensor, respectively. The auxiliary four-vector $\Pi^\mu$,
required for the overall consistency, has the form $\Pi^\mu=2u_\nu F^{[\mu}_{~\,\alpha} M^{\nu]\alpha}$ and satisfies $\Pi^{[\mu} u^{\nu]} = - F^{[\mu}_{~~\alpha} M^{\nu]\alpha}$ \cite{Israel:1978up}, where $X^{[\mu\nu]}\equiv(X^{\mu\nu}-X^{\nu\mu})/2$ denotes antisymmetrization. Note that because of the latter condition, $T^{\mu\nu}_{\mathrm{int }}$ is symmetric in $\mu$ and $\nu$.

In the current work, we are interested in the formulation of non-resistive magnetohydrodynamics with spin. The first term on the right hand side of Eq.~\eqref{T^mn_int1} can be shown to vanish in the limit of infinite conductivity \cite{Florkowski:2018ubm}. In this case, we have
\begin{align}
    T^{\mu\nu}_{\mathrm{int}} &= - F^{\mu}_{~\,\alpha} M^{\nu\alpha}, \label{T^mn_int2}
\end{align}
where the anti-symmetric part of the right hand side vanishes in non-resistive case \cite{Florkowski:2018ubm}. In this case, the field strength tensor is given by $F^{\mu\nu}=\epsilon^{\mu\nu\alpha\beta}u_\alpha B_\beta$, where $B^\mu$ is the magnetic field four-vector. In our metric convention, the field strength tensor and polarization-magnetization tensors are related to each other as\footnote{We note that while this differs from the commonly used $H^{\mu\nu} = F^{\mu\nu} - M^{\mu\nu}$, our choice is appropriate for the kinetic theory definition of $M^{\mu\nu}$, Eq.~\eqref{mag-pol-kin}. } $H^{\mu\nu} = F^{\mu\nu} + M^{\mu\nu}$ \cite{Balakin:2007rp, Balakin:2005fu}, where $H^{\mu\nu}$ is the induction tensor. Taking the divergence of Eq.~\eqref{T^mn_em} one can easily show that,
\begin{align}
    \partial_\nu T^{\mu\nu}_{\mathrm{em}} &= F^{\mu}_{~\,\alpha} J^{\alpha}, \label{div_n T^mn_em}
\end{align}
where we used the Maxwell's equation, which in case of magnetizable medium is given by,
\begin{align}
    \partial_\mu H^{\mu\nu} &= J^\nu. \label{Maxwell's eqn}
\end{align}
Here $J^\mu$ is the charge four-current that generates the electromagnetic field.

It is important to note that $J^\mu$ can have two origins - the charged currents within the fluid system and a current acting as the generator of a background external electromagnetic field i.e. $J^\mu = J^\mu_{\mathrm{f}} + J^\mu_{\rm ext}$ \cite{Denicol:2018rbw, Denicol:2019iyh}. Note that for charged fluid, we have $J^\mu_{\rm f}=\mathfrak{q}N^\mu$, where $\mathfrak{q}$ is the electric charge of the particles. Consequently, the field strength tensor, $F^{\mu\nu}$ is composed of a field due to the charged currents within the fluid and a background external field, $F^{\mu\nu} = F^{\mu\nu}_{\mathrm{f}} + F^{\mu\nu}_{\rm ext}$. In case with non-zero $F^{\mu\nu}_{\rm ext}$, the $T^{\mu\nu}$ expressed in Eq.~\eqref{T^mn_decomp} is not conserved because the energy-momentum contribution of the current, $J^\mu_{\rm ext}$, which produces the external field, is not considered. In this case, the divergence of energy-momentum tensor is equal to a force that we will be calling `external force',
\begin{align}
    \partial_\nu T^{\mu\nu} &= - f^\mu_{\mathrm{ext}}. \label{div_n T^mn_tot}
\end{align}
where, $f^\mu_{\mathrm{ext}} = F^{\mu}_{~\,\alpha} J^{\alpha}_{\mathrm{ext}}$. Using Eqs.~\eqref{T^mn_decomp}, \eqref{T^mn_em}, \eqref{T^mn_int2} and \eqref{div_n T^mn_tot}, the expression for divergence of $T^{\mu\nu}_{\mathrm{f}}$ is found to be,
\begin{align}
    \partial_\nu T^{\mu\nu}_{\mathrm{f}} = F^{\mu}_{~\,\alpha} J^\alpha_{\mathrm{f}} + \frac{1}{2} \left( \partial^\mu F^{\nu\alpha} \right) M_{\nu\alpha}. \label{div_n T^mn_f}
\end{align}
Later, we will prove this relation can be obtained exactly using relativistic Boltzmann equation.

Next, we consider angular momentum conservation. The total angular momentum is the sum of the orbital and spin angular momentum. We write this as
\begin{align}
    J^{\lambda,\mu\nu} &= L^{\lambda,\mu\mu} + S^{\lambda,\mu\nu}, \label{NCL:J^lmn_decomp}
\end{align}
where $L^{\lambda,\mu\nu}$ is the orbital part and $S^{\lambda,\mu\nu}$ is the spin part of the total angular momentum, respectively. In presence of external torque, the total angular momentum is not conserved and its divergence leads to,
\begin{align}
    \partial_\lambda J^{\lambda,\mu\nu} &= - \tau_{\mathrm{ext}}^{\mu\nu}. \label{NCL:tam_consv}
\end{align}
Here we consider the field to be devoid of pure torque and $\tau^{\mu\nu}_\mathrm {ext}$ in the above equation is due to the moment of the external force, i.e.,
\begin{align}
    \tau_{\mathrm{ext}}^{\mu\nu} &= x^\mu f_{\mathrm{ext}}^\nu - x^\nu f_{\mathrm{ext}}^\mu \label{torque_ext}
\end{align}
The orbital part of the total  angular momentum is defined as,
\begin{align} 
    L^{\lambda,\mu\nu} &= x^{\mu} T^{\lambda\nu} - x^{\nu} T^{\lambda\mu}. \label{NCL:L^lmn_decomp}
\end{align}
The divergence of the above equation leads to,
\begin{equation} \label{NCL:oam_cons}
    \partial_\lambda L^{\lambda,\mu\nu} = - x^{\mu} f_{\mathrm{ext}}^\nu + x^{\nu} f_{\mathrm{ext}}^\mu = -\tau_{\mathrm{ext}}^{\mu\nu}.
\end{equation}
Therefore, from Eqs.~\eqref{NCL:J^lmn_decomp}, \eqref{NCL:tam_consv} and \eqref{NCL:oam_cons}, we conclude that the spin part of the total angular momentum is conserved, i.e.,
\begin{equation} \label{NCL:sam_cons}
    \partial_\lambda S^{\lambda,\mu\nu} = 0.
\end{equation}
This seems reasonable because, in the present work, we do not consider the field to carry pure torque which could have affected the conservation of the spin part of total angular momentum. Subsequently, our hydrodynamic evolution equations comprises of Eqs.~\eqref{N^mu_cons}, \eqref{div_n T^mn_f} and \eqref{NCL:sam_cons}. Next, we show that these equations can also be obtained from kinetic theory.



{\bf \textit{Kinetic theory}}-- The phase-space distribution function of particles with intrinsic angular momentum is given by, $f(x,p,s)$, where $x\equiv x^\mu$ is the space-time four vector, $p\equiv p^\mu$ is the four-momentum of the particles and $s\equiv s^{\mu\nu}$ is the classical analogue of particle spin which we define as the internal angular momentum of the particles \cite{Florkowski:2018fap}. The Boltzmann equation governing the evolution of the distribution function can be written as \cite{suttorp1970covariant, weert1970relativistic, dixon1964covariant, supplement}
\begin{align}
    \left( p^\alpha \dfrac{\partial }{\partial x^\alpha} + m\,\mathcal{F}^\alpha \dfrac{\partial }{\partial p^\alpha} + m\,\mathcal{S}^{\alpha\beta} \dfrac{\partial }{\partial s^{\alpha\beta}} \right) f &= C [\,f\,]\,,\label{NCL:Beq_v1}
\end{align}
and similarly for anti-particles with $f\to\bar{f}$. In the above equation, the particle four-momentum has the components $p^\mu = (E_p, {\bf p})$ with $E_p = \sqrt{m^2 + {\bf p}^2}$ and $m$ denoting particle energy and mass, respectively, and $C [f]$ is the collision kernel. Here, $\mathcal{F}^\alpha=dp^\alpha/d\tau$ ($\tau$ being proper time along the world line) is the force experienced by a particle moving under the influence of electromagnetic field which can lead to change in the four-momentum of the particles and $\mathcal{S}^{\alpha\beta}=ds^{\alpha\beta}/d\tau$ is a pure torque term which can lead to change in the internal angular momentum of the particles. 

Using the Frenkel condition, one can derive the force and torque term in the Boltzmann equation as \cite{dixon1964covariant, suttorp1970covariant, weert1970relativistic, Bailey:1975fe, supplement},
\begin{align}
    \mathcal{F}^\alpha &= \frac{\mathfrak{q}}{m}\,  F^{\alpha\beta} p_\beta + \frac{1}{2} \left( \partial^\alpha F^{\beta\gamma} \right) m_{\beta\gamma}, \label{NCL:Beq_Force}\\
    \mathcal{S}^{\alpha\beta} &= 2\, F^{\gamma[\alpha}\, m_{~~\gamma}^{\beta]} - \frac{2}{m^2} \!\left(\! \chi - \frac{\mathfrak{q}}{m} \!\right)\! F_{\phi\gamma}\, s^{\phi[\alpha}\, p^{\beta]} p^\gamma . \label{NCL:Beq_Spin}
\end{align}
The first term on the right hand side of Eq.~\eqref{NCL:Beq_Force} represents the Lorentz force and the second term is known as the Mathisson force. Here the magnetic dipole moment of particles, $m^{\alpha\beta}$, is proportional to the internal angular momentum, i.e., $m^{\alpha\beta}=\chi s^{\alpha\beta}$, with $\chi$ resembling the gyromagnetic ratio \cite{suttorp1970covariant, Weickgenannt:2019dks}. The expression in Eq.~\eqref{NCL:Beq_Spin} is for the pure torque term arising from interaction of the particle magnetic moment and the electromagnetic field. It is important to note that this torque term, obtained for composite particles, affects only the internal angular momentum of the particles. However, its origin from Wigner-function formalism is not understood, as opposed to the force term \cite{Weickgenannt:2019dks}. Therefore, in the present work, we do not consider the pure torque term and leave its analysis for future work. 

Next, we ascertain that appropriate moments of the Boltzmann equation, Eq.~\eqref{NCL:Beq_v1}, leads to the hydrodynamic equations, i.e., Eqs.~\eqref{N^mu_cons}, \eqref{div_n T^mn_f} and \eqref{NCL:sam_cons}. In terms of the distribution function, the particle current, energy-momentum tensor and the spin current of the fluid can be written as \cite{Bhadury:2020puc}
\begin{eqnarray}
N^\mu &=& \int dP dS \, p^\mu \left(f - \bar{f} \right), \label{eq:classNTS1} \\
T^{\mu\nu}_{\rm f} &=& \int dP dS \, p^\mu p^\nu \left(f + \bar{f} \right), \label{eq:classNTS2} \\
S^{\lambda,\mu\nu} &=& \int dP dS \, p^\lambda s^{\mu\nu} \left(f + \bar{f} \right),
\label{eq:classNTS3}
\end{eqnarray}
where $dP \equiv  d^3p/[E_p (2 \pi )^3]$ and $dS \equiv m/(\pi \spin) \,  d^4s \, \delta(s \cdot s + \spin^2) \, \delta(p \cdot s)$ with the length of the spin vector defined by the eigenvalue of the Casimir operator, $\spin^2 = \frac{1}{2} \left( 1+ \frac{1}{2}  \right) = \frac{3}{4}$. In terms of the distribution function, one can also define the polarization-magnetization tensor as \cite{suttorp1970covariant,weert1970relativistic}
\begin{equation}\label{mag-pol-kin}
M^{\mu\nu} = m\int dP dS \, m^{\mu\nu} \left(f - \bar{f} \right),
\end{equation}
whose equilibrium expression is obtained in Ref.~\cite{supplement}.

Assuming that the microscopic interactions do not violate fundamental conservation laws, we have vanishing zeroth and first moment of the collision kernel, i.e., $\int dP dS\, C [f] = \int dP dS\,p^\mu\, C [f]=0$. We also impose a matching condition for the spin current such that the `spin moment' of the collision kernel vanishes, i.e., $\int dP dS\, s^{\mu\nu}C[f]=0$ \cite{Bhadury:2020puc}. This condition ensures that the collisions preserve internal angular momentum of the particles. Using the definitions of the fluid currents in terms of the distribution function, Eqs.~\eqref{eq:classNTS1}-\eqref{eq:classNTS3}, and properties of the collision kernel as described above, we find that the zeroth, first and spin moment of the Boltzmann equation, Eq.~\eqref{NCL:Beq_v1}, in absence of the torque term, leads to Eqs.~\eqref{N^mu_cons}, \eqref{div_n T^mn_f} and \eqref{NCL:sam_cons}, respectively. This is an important result of the present work which sets up the basis for the formulation of spin-magnetohydrodynamics from kinetic theory.



{\bf \textit{Dissipative hydrodynamics}}-- In order to derive constitutive relations for dissipative quantities, we consider the Boltzmann equation, without the pure torque term, in relaxation-time approximation \cite{anderson1974relativistic},
\begin{align}
    \left( p^\alpha \dfrac{\partial }{\partial x^\alpha} + m\,\mathcal{F}^\alpha \dfrac{\partial }{\partial p^\alpha} \right) f &= -\left(u\cdot p\right)\frac{f-f_{\rm eq}}{\tau_{\rm eq}},\label{NCL:Beq_RTA}
\end{align}
where $u\cdot p\equiv u_\mu p^\mu$, $f_{\rm eq}$ is the equilibrium distribution function and $\tau_{\rm eq}$ is the relaxation time which, in the present work is assumed to be independent of particle momentum/energy. Note that, the collision kernel in the relaxation-time approximation, i.e., right hand side of the above equation, has vanishing zeroth and first moment with Landau frame definition of the fluid velocity. Vanishing of spin moment is guaranteed if we impose the matching condition \cite{Bhadury:2020puc}
 \begin{equation}
   u_\lambda \delta S^{\lambda,\mu\nu} \equiv u_\lambda\left( S^{\lambda,\mu\nu} - S^{\lambda,\mu\nu}_{\rm eq} \right) = 0, \label{eq:LS}  
 \end{equation}
where, $\delta S^{\lambda,\mu\nu}$ is the non-equilibrium correction to the spin current. With the above condition, along with the Landau frame and matching conditions, the zeroth, first and spin moment of Eq.~\eqref{NCL:Beq_RTA} leads to the hydrodynamic equations, Eqs.~\eqref{N^mu_cons}, \eqref{div_n T^mn_f} and \eqref{NCL:sam_cons}, respectively. 

In the present work, we consider equilibrium distribution to be described by Fermi-Dirac statistics,
\begin{equation}\label{feq_FD}
    f_{\rm eq} = \frac{1}{1+\exp\left[\beta (u\!\cdot\! p) - \xi - \frac{1}{2}\,\omega:s \right]},
\end{equation}
where, $\beta\equiv 1/T$ is the inverse temperature, $\xi\equiv \mu/T$ is the ratio of chemical potential and temperature, and $\omega:s\equiv\omega_{\mu\nu}s^{\mu\nu}$. Here, $\omega_{\mu\nu}$ is a Lagrange multiplier corresponding to angular momentum conservation \cite{Florkowski:2017ruc} and is related to spin polarization observable via Pauli-Lubanski four-vector \cite{Florkowski:2017dyn, Florkowski:2018fap}. For anti-particles, one can obtain the equilibrium distribution $\bar{f}_{\rm eq}$ from  the above equation with the replacement $\xi\to-\xi$. In the current formulation, we work in the small polarization limit. Hence, keeping terms up to linear order in $\omega^{\mu\nu}$, one can write the equilibrium distribution function as
\begin{equation} \label{feq_f0}
    f_{\rm eq} = f_0 + \frac{1}{2}\left(\omega:s\right) f_0 \Tilde{f}_0,
\end{equation}
where, $f_0\equiv\left\{ 1+\exp\left[\beta(u\!\cdot\!p) - \xi\right] \right\}^{-1}$ and $\Tilde{f}_0\equiv 1-f_0$.

Using $f=f_{\rm eq}$ and $\bar{f}=\bar{f}_{\rm eq}$ in Eq.~\eqref{mag-pol-kin}, we find that the equilibrium expression for magnetization tensor is linear in $\omega^{\mu\nu}$ and takes the form $M^{\mu\nu}_{\rm eq}=a_1\,\omega^{\mu\nu} + a_2\, u^{[\mu} u_\gamma \omega^{\nu]\gamma}$ \cite{supplement}. In order to make connection with the Barnett effect, we note that, in global equilibrium, $\omega^{\mu\nu}$ corresponds to rotation of the fluid \cite{Becattini:2007nd, Becattini:2009wh, Becattini:2015nva, Florkowski:2018fap, Florkowski:2017ruc, Florkowski:2018ahw, Florkowski:2019qdp, Florkowski:2021wvk, Florkowski:2017dyn}. Therefore, from the expression of $M^{\mu\nu}_{\rm eq}$, we conclude that rotation of the fluid produces magnetization, which is precisely the physics of Barnett effect \cite{Barnett, Hernandez:2017mch}. This expression also implies the converse, i.e., Einstein-de Haas effect.

The expressions for dissipative quantities that we need to obtain are $n^\mu$ defined in Eq.~\eqref{Nmu}, $\Pi$ and $\pi^{\mu\nu}$ defined in Eq.~\eqref{T^mn_f} and $\delta S^{\lambda,\mu\nu}$ defined in Eq.~\eqref{eq:LS}. In terms of the non-equilibrium corrections to the distribution function, $\delta f=f-f_{\rm eq}$ and $\delta\bar{f}=\bar{f}-\bar{f}_{\rm eq}$, these dissipative quantities can be expressed as
\begin{eqnarray}
    n^\mu &=& \Delta^\mu_\alpha \int dP dS \, p^\alpha \left(\delta f - \delta\bar{f} \right), \label{diss1} \\
    \Pi &=& -\frac{1}{3}\Delta_{\alpha\beta} \int dP dS \, p^\alpha p^\beta \left(\delta f + \delta\bar{f} \right), \label{diss2} \\
    \pi^{\mu\nu} &=& \Delta^{\mu\nu}_{\alpha\beta} \int dP dS \, p^\alpha p^\beta \left(\delta f + \delta\bar{f} \right), \label{diss3} \\
    \delta S^{\lambda,\mu\nu} &=& \int dP dS \, p^\lambda s^{\mu\nu} \left(\delta f + \delta\bar{f} \right), \label{diss4}
\end{eqnarray}
where $\Delta^{\mu\nu}_{\alpha\beta}\equiv \frac{1}{2}(\Delta^{\mu}_{\alpha}\Delta^{\nu}_{\beta} + \Delta^{\mu}_{\beta}\Delta^{\nu}_{\alpha}) - \frac{1}{3}\Delta^{\mu\nu}\Delta_{\alpha\beta}$ is a traceless symmetric projection operator which is orthogonal to both $u^\mu$ and $\Delta^{\mu\nu}$. To obtain the relativistic Navier-Stokes expressions for the above dissipative quantities, we need to evaluate $\delta f$ and $\delta\bar{f}$ up to first-order in hydrodynamic gradients. To that end, we employ the Boltzmann equation in relaxation-time approximation, Eq.~\eqref{NCL:Beq_RTA}.

Using the matching condition, Eq.~\eqref{eq:LS}, we obtain the evolution equation for the spin polarization tensor,
\begin{align}
    &\Dot{\omega}^{\mu\nu} \!=\! \mathcal{D}_{\Pi}^{[\mu\nu]}\, \theta 
    +\! \mathcal{D}_{a}^{[\mu\nu]\gamma} \Dot{u}_\gamma 
    +\! \mathcal{D}_{\mathrm{n}}^{[\mu\nu]\gamma} \left( \nabla_\gamma \xi \right) 
    +\! \mathcal{D}_{B}^{[\mu\nu]\rho\kappa}\! \left( \nabla_{\rho} B_{\kappa} \right) \nonumber\\
    &+ \mathcal{D}_{\pi}^{[\mu\nu]\rho\kappa} \sigma_{\rho\kappa} 
    + \mathcal{D}_{\Omega}^{[\mu\nu]\rho\kappa} \Omega_{\rho\kappa}\! 
    +\! \mathcal{D}_{\Sigma}^{[\mu\nu]\phi\rho\kappa}\! \left( \nabla_\phi \omega_{\rho\kappa} \right)\!, \label{omegadot}
\end{align}
where $\dot{X}\equiv u^\alpha\partial_\alpha X$, $\nabla^\mu \equiv\Delta^{\mu\alpha}\partial_\alpha$, $\sigma^{\mu\nu}\equiv\Delta^{\mu\nu}_{\alpha\beta} (\partial^\alpha u^\beta)$ and $\Omega_{\mu\nu} \equiv (\partial_\mu u_\nu - \partial_\nu u_\mu )/2$ is the fluid vorticity tensor. In the above equation dependence on hydrodynamic gradients is made explicit and the tensor coefficients, $\mathcal{D}$, contains equilibrium quantities. The explicit forms of these coefficients are provided in Ref.~\cite{supplement}. We see that the above equation contains information about the connection between evolution of spin polarization tensor, $\omega^{\mu\nu}$, and fluid vorticity, $\Omega_{\mu\nu}$, via the term having coefficient $\mathcal{D}_{\Omega}^{\mu\nu\rho\kappa}$. It is important to note that the coefficient $\mathcal{D}_{\Omega}^{\mu\nu\rho\kappa}$ vanishes in absence of electromagnetic field which leads us to conclude that the conversion of spin-polarization to thermal vorticity proceeds via coupling with electromagnetic field. While vorticity terms have previously been obtained in the constitutive equations in absence of electromagnetic field \cite{Baier:2007ix, Romatschke:2009kr, Moore:2010bu}, we obtain here, for the first time, such coupling terms. Another important feature of Eq.~\eqref{omegadot} is that the coupling between magnetic field and spin polarization occur at gradient order. 

The non-equilibrium correction to the phase-space distribution function, up to first-order in space-time gradients, is obtained from Eq.~\eqref{NCL:Beq_RTA} as
\begin{widetext}
\begin{equation}\label{KT:delf1}
    \delta f_1 = - \frac{\tau_{\mathrm{R}}}{\left(u\!\cdot\! p\right)} \!\left[ p^\alpha \partial_\alpha + \frac{m\, \chi}{2} \left( \partial^\alpha F^{\beta\gamma} \right) s_{\beta\gamma} \partial^{(p)}_\alpha \right]\! f_{\rm eq} + \frac{\tau_{\mathrm{R}}}{\left(u\!\cdot\! p\right)} \mathfrak{q} F^{\alpha\beta} p_\beta \partial^{(p)}_\alpha \!\left[ \frac{\tau_{\mathrm{R}}}{\left(u\!\cdot\! p\right)} \!\left\{ p^\rho \partial_\rho + \frac{m\, \chi}{2} \left( \partial^\rho F^{\phi\kappa} \right) s_{\phi\kappa} \partial^{(p)}_\rho \right\}\! f_{\rm eq} \right], 
\end{equation}
where, $\partial^{(p)}_\alpha\equiv \frac{\partial}{\partial p^\alpha}$ is the partial derivative with respect to particle momenta. To obtain first-order non-equilibrium correction for anti-particles, $\delta\bar{f}_1$, one has to replace $f\to \Bar{f}$, $\xi \to - \xi$ and, $\mathfrak{q}\to - \mathfrak{q}$ in the above equation. Substituting the first-order non-equilibrium corrections, $\delta f_1$ and $\delta\bar{f}_1$ in Eqs.~\eqref{diss1}-\eqref{diss4}, we obtain the constitutive relations for the dissipative quantities as,
\begin{align}
    n^{\mu} &= \tau_{\mathrm{eq}} \left[ \beta_{n\Pi}^{\langle\mu\rangle}\, \theta 
    + \beta_{na}^{\langle\mu\rangle\alpha} \dot{u}_{\alpha}
    + \beta_{nn}^{\langle\mu\rangle\alpha} \left( \nabla_\alpha\xi \right) 
    + \beta_{nF}^{\langle\mu\rangle\alpha\beta} \left( \nabla_{\alpha} B_{\beta} \right) 
    + \beta_{n\pi}^{\langle\mu\rangle\alpha\beta} \sigma_{\alpha\beta}
    + \beta_{n\Omega}^{\langle\mu\rangle\alpha\beta} \Omega_{\alpha\beta} 
    + \beta_{n\Sigma}^{\langle\mu\rangle\alpha\beta\gamma} \left( \nabla_\alpha \omega_{\beta\gamma} \right) \right], \label{nmu_final}\\
    \Pi &= \tau_{\mathrm{eq}} \left[ \beta_{\Pi\Pi}\, \theta 
    + \beta_{\Pi a}^{\alpha} \dot{u}_{\alpha} 
    + \beta_{\Pi n}^{\alpha} \left( \nabla_\alpha\xi \right) 
    + \beta_{\Pi F}^{\alpha\beta} \left( \nabla_{\alpha} B_{\beta} \right) 
    + \beta_{\Pi\pi}^{\alpha\beta} \sigma_{\alpha\beta} 
    + \beta_{\Pi\Omega}^{\alpha\beta} \Omega_{\alpha\beta} 
    + \beta_{\Pi\Sigma}^{\alpha\beta\gamma} \left( \nabla_\alpha \omega_{\beta\gamma} \right) \right], \label{Pi_final}\\
    \pi^{\mu\nu} &= \tau_{\mathrm{eq}} \left[ \beta_{\pi\Pi}^{\langle\mu\nu\rangle} \theta 
    \!+ \beta_{\pi a}^{\langle\mu\nu\rangle\alpha} \dot{u}_{\alpha} 
    \!+ \beta_{\pi n}^{\langle\mu\nu\rangle\alpha}\! \left(\! \nabla_\alpha\xi \right)
    \!+\! \beta_{\pi F}^{\langle\mu\nu\rangle\alpha\beta}\! \left(\! \nabla_{\alpha} B_{\beta}\! \right) 
    \!+\! \beta_{\pi\pi}^{\langle\mu\nu\rangle\alpha\beta} \sigma_{\alpha\beta} 
    \!+\! \beta_{\pi\Omega}^{\langle\mu\nu\rangle\alpha\beta} \Omega_{\alpha\beta}
    \!+\! \beta_{\pi\Sigma}^{\langle\mu\nu\rangle\alpha\beta\gamma}\! \left(\! \nabla_\alpha \omega_{\beta\gamma} \right) \right]\!, \label{pimunu_final}\\
    \delta S^{\lambda,\mu\nu} &= \tau_{\mathrm{eq}}\, \Big[ B_{\Pi}^{\lambda,[\mu\nu]}\, \theta 
    + B_{a}^{\lambda,[\mu\nu]\alpha} \dot{u}_{\alpha} 
    + B_{n}^{\lambda,[\mu\nu]\alpha} \!\left( \nabla_\alpha \xi \right) 
    + B_{F}^{\lambda,[\mu\nu]\alpha\beta}\! \left( \nabla_{\alpha} B_{\beta} \right) 
    + B_{\pi}^{\lambda,[\mu\nu]\alpha\beta} \sigma_{\alpha\beta} 
    + B_{\Omega}^{\lambda,[\mu\nu]\alpha\beta} \Omega_{\alpha\beta} \nonumber\\
    &\qquad\quad+ B_{\Sigma}^{\lambda,[\mu\nu]\alpha\beta\gamma}\! \left( \nabla_\alpha \omega_{\beta\gamma} \right) \Big], \label{deltaS_final}
\end{align}
\end{widetext}
where, $X^{\langle\mu\rangle}\equiv\Delta^{\mu}_{\alpha}X^\alpha$ represents projection of a vector orthogonal to fluid four-velocity and $X^{\langle\mu\nu\rangle}\equiv\Delta^{\mu\nu}_{\alpha\beta}X^{\alpha\beta}$ denotes traceless symmetric projection of a two-rank tensor. The above equations represent the first result of relativistic formulation of spin-magnetohydrodynamics. We find that all dissipative quantities are affected by several hydrodynamic gradients and contain coupling between spin and magnetic field. The detailed expressions for the tensor transport coefficients, appearing in the above equation, are provided in Ref.~\cite{supplement}. Very interestingly, we observe that apart from usual hydrodynamic gradients, Eqs.~\eqref{nmu_final}-\eqref{deltaS_final} also contain gradients of magnetic field.

In order to identify which first-order gradient terms, appearing in Eqs.~\eqref{nmu_final}-\eqref{deltaS_final}, are dissipative, it is important to compute entropy production in the system. We consider the entropy four-current from the Boltzmann H-theorem,
\begin{equation}\label{Smu}
{\cal H}^\mu = - \!\!\int\! \mathrm{dP} \mathrm{dS}\, p^\mu \left[ \left( f \ln f + \Tilde{f} \ln\Tilde{f} \right) \!+\! \left( \Bar{f} \ln \Bar{f} + \Tilde{\Bar{f}} \ln\Tilde{\Bar{f}} \right) \right].
\end{equation}
Demanding that the divergence of the above entropy current is positive definite, i.e., $\partial_\mu{\cal H}^\mu\geq 0$, we obtain \cite{supplement}
\begin{align}
\Pi = - \zeta \theta, \quad n^{\mu} &= \kappa^{\mu\alpha} \left( \nabla_{\alpha} \xi \right), \quad \pi^{\mu\nu} = \eta^{\mu\nu\alpha\beta} \sigma_{\alpha\beta}, \label{Smu_diss1}\\
\delta S^{\mu,\alpha\beta} &= \Sigma^{\mu\alpha\beta\lambda\gamma\rho} \left(\nabla_\lambda \omega_{\gamma\rho} \right). \label{Smu_diss2}
\end{align}
From the above analysis, we conclude that only those first-order gradient terms which appear in Eqs.~\eqref{Smu_diss1}-\eqref{Smu_diss2}, are dissipative in nature. Comparing Eqs.~\eqref{nmu_final}-\eqref{deltaS_final} and Eqs.~\eqref{Smu_diss1}-\eqref{Smu_diss2}, we obtain $\zeta=-\tau_{\mathrm{eq}} \beta_{\Pi\Pi}$, $\kappa^{\mu\alpha}=\tau_{\mathrm{eq}}\beta_{nn}^{\langle\mu\rangle\alpha}$, $\eta^{\mu\nu\alpha\beta}=\tau_{\mathrm{eq}}\beta_{\pi\pi}^{\langle\mu\nu\rangle\alpha\beta}$ and $\Sigma^{\lambda\mu\nu\alpha\beta\gamma}=\tau_{\mathrm{eq}}B_{\Sigma}^{\lambda,[\mu\nu]\alpha\beta\gamma}$. It is important to note that these dissipative transport coefficients contain coupling between magnetic field and spin polarization/magnetization tensor \cite{supplement}.



{\bf \textit{Summary and Outlook}}-- We presented the first formulation of relativistic spin-magnetohydrodynamics within the kinetic theory framework for spin-$\nicefrac{1}{2}$ particles. We derived equations for relativistic dissipative non-resistive magnetohydrodynamics in the limit of small polarization. We used a relaxation time approximation for the collision kernel in the relativistic Boltzmann equation and calculated non-equilibrium corrections to the phase-space distribution function of spin-polarizable particles. We demonstrated that multiple transport coefficients, dissipative as well as non-dissipative, are present for such a system. We showed that our framework naturally leads to the emergence of the well known Einstein-de Hass and Barnett effects. Further, our results also show that the coupling between the magnetic field and spin polarization appears at gradient order.

Looking forward, it will be interesting to consider a generalization of the above framework to include resistive effects to the flow of charge current. Given that several gradients are present in all dissipative currents, the present first-order theory may prove to be causal and stable, even though it is formulated in the Landau frame. Therefore it is important to perform a stability analysis \cite{Biswas:2020rps, Ambrus:2022yzz, Hu:2022lpi} of Eqs.~\eqref{nmu_final}-\eqref{deltaS_final}. Nonetheless, the present framework can also be extended to include second-order gradients in order to formulate second-order spin-magnetohydrodynamics. We leave these problems for future work.

Finally, we would like to outline another important implication of our formulation in the context of relativistic heavy-ion collisions. Global polarization of $\Lambda$ is generally attributed to large angular momentum generated in non-central collisions. On the other hand, it has been observed that at low energy collisions, there is a noticeable difference of $\Lambda$ and anti-$\Lambda$ polarization \cite{STAR:2017ckg} which can not be explained by global angular momentum alone. It was conjectured that the coupling between magnetic field and intrinsic magnetic moment of emitted particles may induce a larger polarization for anti-$\Lambda$ compared to $\Lambda$ \cite{STAR:2017ckg, Becattini:2016gvu}. Therefore, a simulation based on our unified framework of spin-magnetohydrodynamics has the potential to explain this difference of $\Lambda$ and anti-$\Lambda$ polarization, which we leave for future work.


\begin{acknowledgements}
A.J. was supported in part by the DST-INSPIRE faculty award under Grant No. DST/INSPIRE/04/2017/000038. R.R. was supported in part by the Polish National Science Centre Grants No. 2018/30/E/ST2/00432.
\end{acknowledgements}


\bibliography{ref}



\clearpage

\widetext
\begin{center}
\underline{\textbf{\Large Supplemental Material: Relativistic spin-magnetohydrodynamics}}
\end{center}
\setcounter{equation}{0}
\setcounter{figure}{0}
\setcounter{table}{0}
\setcounter{page}{1}
\makeatletter
\renewcommand{\theequation}{S\arabic{equation}}
\renewcommand{\thefigure}{S\arabic{figure}}

\vspace{0.1cm}
Here we provide brief details of derivation of the Boltzmann equation as well as the equilibrium magnetization tensor, and list the various transport coefficients arising in spin-magnetohydrodynamics as a consequence of coupling between particle spin and the magnetic field arising due to the non-zero magnetization. We also provide details of the derivation of entropy production and determine which gradients are dissipative in nature. These details will help interested readers to understand the calculations and further phenomenological development of the theory.

\vspace{0.2cm}
\begin{center}
\underline{\textbf{Derivations of the Boltzmann equation, its associated force and pure torque terms}}
\end{center}

In the present case, the distribution function of the extended phase-space i.e. $f(x,p,s)$ is function of spacetime, $x^{\mu}$, momentum four-vector, $p^{\mu}$, and internal angular momentum, $s^{\mu\nu}$, of the particles. Moreover, these quantities $x^{\mu}$, $p^{\mu}$, and $s^{\mu\nu}$, are all functions of an affine parameter $\tau$ along the world lines of the particles. Hence, following Liouville's theorem, in the collision-less limit we can write \cite{weert1970relativistic},
\begin{align}
    \frac{d}{d \tau} f(x,p,s) &\equiv \frac{\partial f}{\partial x^\alpha} \frac{dx^\alpha}{d\tau} + \frac{\partial f}{\partial p^{\alpha}} \frac{dp^\alpha}{d\tau} + \frac{\partial f}{\partial s^{\alpha\beta}} \frac{ds^{\alpha\beta}}{d\tau} = 0 \label{Liouville 1}
\end{align}
Next we note that,
\begin{align}
    \frac{d x^\alpha}{d\tau} = \frac{\partial x^\alpha}{\partial t} \frac{dt}{d\tau} = \left( \frac{\partial t}{\partial t} ,\, \frac{\partial x^i}{\partial t}\right)\! \gamma = \Big( 1 ,\, v^i \Big) \gamma = \left( \frac{p^0}{p^0} ,\, \frac{p^i}{p^0} \right) \gamma = p^\alpha \frac{\gamma}{p^0} = \frac{p^\alpha}{m} \label{intermediate 1}
\end{align}
where, $\gamma$ is the Lorentz factor, $m$ is the particle mass and $i$ takes values $(1,2,3)$. Furthermore, we define,
\begin{align}
    \mathcal{F}^\alpha \equiv \frac{d p^\alpha}{d \tau} \qquad \qquad \mathrm{and,} \qquad \qquad \mathcal{S}^{\alpha\beta} \equiv \frac{d s^{\alpha\beta}}{d \tau} \label{intermediate 2n3}.
\end{align}
Multiplying `$m$' on both sides of Eq.~\eqref{Liouville 1} and using the above definitions, we get,
\begin{align}
    p^\alpha \frac{\partial f}{\partial x^\alpha} + m \mathcal{F}^\alpha \frac{\partial f}{\partial p^{\alpha}} + m \mathcal{S}^{\alpha\beta} \frac{\partial f}{\partial s^{\alpha\beta}} = 0 \label{Liouville 2}
\end{align}
In presence of collisions, the right hand side of Eq.~\eqref{Liouville 2} is not zero and has to be replaced by a collision kernel as,
\begin{align}
    p^\alpha \frac{\partial f}{\partial x^\alpha} + m \mathcal{F}^\alpha \frac{\partial f}{\partial p^{\alpha}} + m \mathcal{S}^{\alpha\beta} \frac{\partial f}{\partial s^{\alpha\beta}} = C[f]. \label{Beq_sup}
\end{align}
Above equation is the Boltzmann equation governing the evolution of the distribution function with extended phase-space, Eq.~(18), of the main text.  

The next step is to obtain the explicit expressions of the force and pure torque terms of the Boltzmann equation. For this purpose, we follow Ref. \cite{suttorp1970covariant}, where starting from the (Mathisson-Papapetrau-Dixon) MPD-like  equations for electromagnetic fields \cite{Bailey:1975fe}, one arrives at the following expressions of the equations of motion and spin as,
\begin{align}
    \frac{d p^\alpha}{d\tau} &= \mathfrak{q}\,  F^{\alpha\beta}\, \mathcal{U}_\beta + \frac{1}{2} \left(\partial^\alpha F^{\beta\gamma} \right) m_{\beta\gamma} \label{eom 1}\\
    \frac{d s^{\alpha\beta}}{d \tau} &= p^\alpha\, \mathcal{U}^\beta - p^\beta\, \mathcal{U}^\alpha + F^{\alpha\gamma}\, m_\gamma^{~\beta} - F^{\beta\gamma}\, m_\gamma^{~\alpha} \label{eos 1}
\end{align}
where, $\mathcal{U}^\alpha$ is the four-velocity of individual particles and not the fluid four-velocity. It is important to note that, $p^\mu$ and $\mathcal{U}^\mu$ are not parallel for the spin-\nicefrac{1}{2} particles in presence of magnetic field and thus follow the relation \cite{suttorp1970covariant},
\begin{align}
    p^\beta = m\, \mathcal{U}^\beta + \frac{1}{2\, m} F_{\alpha\gamma} m^{\alpha\gamma}\, p^\beta + \frac{\mathfrak{q}}{m^2} F_{\alpha\gamma}\, p^\gamma\, s^{\alpha\beta} + \frac{1}{m} p^\alpha\, F_{\alpha\gamma} m^{\gamma\beta}. \label{mom-vel reln}
\end{align}
Note that the above equation was derived using the Frenkel condition, $p^\alpha s_{\alpha\beta} = 0$. We differentiated the Frenkel condition, used Eqs.~\eqref{eom 1} and \eqref{eos 1} and ignored higher order terms. Finally, substituting the particle four-velocity from Eq.~\eqref{mom-vel reln} into Eqs.~\eqref{eom 1} and \eqref{eos 1} and dropping terms that would lead to terms higher than quadratic in magnetic fields, we can obtain the equations of motion and spin in terms of kinetic theory variables as,
\begin{align}
    \mathcal{F}^\alpha &\equiv \frac{d p^\alpha}{d\tau} = \frac{\mathfrak{q}}{m}\,  F^{\alpha\beta} p_\beta + \frac{1}{2} \left( \partial^\alpha F^{\beta\gamma} \right) m_{\beta\gamma}, \label{eom 2}\\
    \mathcal{S}^{\alpha\beta} &\equiv \frac{d s^{\alpha\beta}}{d \tau} = 2\, F^{\gamma[\alpha}\, m_{~~\gamma}^{\beta]} - \frac{2}{m^2} \!\left(\! \chi - \frac{\mathfrak{q}}{m} \!\right)\! F_{\phi\gamma}\, s^{\phi[\alpha}\, p^{\beta]} p^\gamma . \label{eos 2}
\end{align}
Thus, we find that the Eqs.~\eqref{Beq_sup}, \eqref{eom 2} and \eqref{eos 2} are exactly the same as those given in Eqs.~(18), (19) and (20) of the main text, respectively.

\vspace{0.2cm}
\begin{center}
\underline{\textbf{Equilibrium polarization-magnetization tensor}}
\end{center}
\vspace{0.1cm}

One may use the definition of the polarization-magnetization tensor given in Eq.~(24), given in the main text, to obtain its equilibrium expression, $M^{\alpha\beta}_{\rm eq}$, as
\begin{align}
    M^{\alpha\beta}_{\rm eq} = m \int dP dS \, m^{\mu\nu} \left(f_{\rm eq} - \bar{f}_{\rm eq} \right)
    = a_1\,\omega^{\alpha\beta} + a_2\, u^{[\alpha} u_\gamma \omega^{\beta]\gamma},
\end{align}
where,
\begin{equation}
a_1 = \frac{2 \chi \mathfrak{s}^2}{3 m^2} \left( m^2 J^{-}_{00} - 2  J^{-}_{21} \right), \qquad a_2 = \frac{4 \chi \mathfrak{s}^2}{3 m^2}  \left( J^{-}_{20} -  J^{-}_{21} \right).
\end{equation}

We define the thermodynamic integrals appearing in this supplemental material as,
\begin{align}
    &I_{n,q}^{\pm} \equiv \frac{1}{(2q+1)!!} \!\int \!dP \left( u \!\cdot\! p \right)^{\!n-2q} \!\left( \Delta_{\alpha\beta}\, p^\alpha p^\beta \right)^{\!q} \! \left(f_{0} \pm \bar{f}_{0} \right), \label{I_nq} \\
    &I^{\mu_1 \mu_2 \cdots \mu_n}_{(r)\pm} \equiv \!\int \!\frac{dP}{\left( u \!\cdot\! p \right)^{r}} p^{\mu_1} p^{\mu_2} \cdots p^{\mu_n} \left(f_{0} \pm \bar{f}_{0} \right). \label{I^mn}
\end{align}
The $J$-type and $K$-type thermodynamic integrals are defined by replacing $\left(f_{0} \pm \Bar{f}_{0} \right)$ in the above equations with $\left(f_{0} \Tilde{f}_0 \pm \Bar{f}_{0} \Tilde{\Bar{f}}_0 \right)$ and $\left(f_{0} \Tilde{f}_0 \hat{f}_0 \pm \Bar{f}_{0} \Tilde{\Bar{f}}_0 \hat{\Bar{f}}_0 \right)$, respectively, where $\hat{f} = 1 - 2 f_0$.

\vspace{0.2cm}
\begin{center}
\underline{\textbf{Expressions for the transport coefficients}}
\end{center}

We start by listing the $\mathcal{D}$-type coefficients, appearing in Eq.~(33) of the main text, as
\begin{align}
    \mathcal{D}_{\Pi}^{\mu\nu} &= \tau_{\mathrm{eq}} \mathcal{D}_{\mathrm{n}2,\,\gamma}^{\mu\nu} \beta_{n\Pi}^{\langle\gamma\rangle} + \left( G_{J,\,10}^{+} \right)^{-1} \bigg[ \beta \left\{ \left( m^2 K^{+}_{21} - \frac{10}{3} K^{+}_{42} \right) \omega^{\mu\nu} + 2 \left( K^{+}_{41} - \frac{5}{3} K^{+}_{42} \right) u^{[\mu} \left( u_\alpha \omega^{\nu]\alpha} \right) \right\} \nonumber\\
    &~~~+ \beta_\theta \left\{ G_{K,\,20}^{+} \omega^{\mu\nu} + 2 \widetilde{G}_{K,\,40}^{+} u^{[\mu} \left( u_\alpha \omega^{\nu]\alpha} \right) \right\} - \xi_\theta \left\{ G_{K,\,10}^{-} \omega^{\mu\nu} + 2 \widetilde{G}_{K,\,30}^{-} u^{[\mu} \left( u_\alpha \omega^{\nu]\alpha} \right) \!\right\} - 2 \widetilde{G}_{J,\,30}^{+} u^{[\mu} \mathcal{C}_{\Pi}^{\nu]} \bigg] \label{D_Pi} \\
    \mathcal{D}_{\mathrm{a}}^{\mu\nu\gamma} &= \tau_{\mathrm{eq}} \mathcal{D}_{\mathrm{n}2,\,\rho}^{\mu\nu} \beta_{na}^{\langle\rho\rangle\gamma} \label{D_a} \\
    \mathcal{D}_{\mathrm{n}}^{\mu\nu\gamma} &= \left( G_{J,\,10}^{+} \right)^{-1} \bigg[ - 2 K^{-}_{31} \left\{ u_\rho g^{\gamma[\mu} \omega^{\nu]\rho} + u^{[\mu}\omega^{\nu]\gamma} \right\} + 2 K_{41}^{+} \left\{ \frac{n_{\rm f}}{\left( \epsilon + P \right)} g^{\gamma[\mu} \left( u_\rho \omega^{\nu]\rho} \right) + \frac{n_{\rm f}}{\left( \epsilon + P \right)} u^{[\mu} \omega^{\nu]\gamma} \right\} \nonumber\\
    &~~~- 2 \widetilde{G}_{J,\,30}^{+} u^{[\mu} \mathcal{C}^{\nu]\gamma}_{\mathrm{n}1} \bigg] + \tau_{\mathrm{eq}} \mathcal{D}_{\mathrm{n}2,\,\rho}^{\mu\nu} \beta_{nn}^{\langle\rho\rangle\gamma} \label{D_n} \\
    \mathcal{D}_{\mathrm{F}}^{\mu\nu\rho\kappa} &= \tau_{\mathrm{eq}} \mathcal{D}_{\mathrm{n}2,\,\gamma}^{\mu\nu}    \beta_{nF}^{\langle\gamma\rangle\rho\kappa} \label{D_F} \\
    \mathcal{D}_{\pi}^{\mu\nu\rho\kappa} &= - 2 \left( G_{J,\,10}^{+} \right)^{\!-1} \!\left( 2 \beta K^{+}_{42} \omega^{\rho[\mu} g^{\nu]\kappa} + \widetilde{G}_{J,\,30}^{+} u^{[\mu} \mathcal{C}^{\nu]\rho\kappa}_{\pi} \right) + \tau_{\mathrm{eq}} \mathcal{D}_{\mathrm{n}2,\,\gamma}^{\mu\nu} \beta_{n\pi}^{\langle\gamma\rangle\rho\kappa} \label{D_pi} \\
    \mathcal{D}_{\Omega}^{\mu\nu\rho\kappa} &= \tau_{\mathrm{eq}} \mathcal{D}_{\mathrm{n}2,\,\gamma}^{\mu\nu} \beta_{n\Omega}^{\langle\gamma\rangle\rho\kappa} \label{D_Omega} \\
    \mathcal{D}_{\Sigma}^{\mu\nu\phi\rho\kappa} &= - 2 \left( G_{J,\,10}^{+} \right)^{-1} \bigg[ J_{31}^{+} \left( u^\kappa g^{\phi[\mu} g^{\nu]\rho} + u^{[\mu} g^{\nu]\rho} g^{\phi\kappa} \right) + \widetilde{G}_{J,\,30}^{+} u^{[\mu} \mathcal{C}^{\nu]\rho}_{\Sigma} g^{\phi\kappa} \bigg] + \tau_{\mathrm{eq}} \mathcal{D}_{\mathrm{n}2,\,\gamma}^{\mu\nu} \beta_{n\Sigma}^{\langle\gamma\rangle\phi\rho\kappa} \label{D_Sigma}
\end{align}
where,
\begin{align}
    G_{i,\,nq}^{\pm} &= m^2\, i_{nq}^{\pm} - 2\, i_{n+2,q+1}^{\pm} \label{DH:abrv1}\\
    \widetilde{G}_{i,\,nq}^{\pm} &= i_{nq}^{\pm} - i_{n,q+1}^{\pm} \label{DH:abrv2}\\
    \mathcal{G}_{i,\,nq}^{\pm} &= G_{i,\,nq}^{\pm} - \widetilde{G}_{i,\,n+2,q}^{\pm} \label{DH:abrv3}
\end{align}
with $i = I,J,K$. In the above expressions of the transport coefficients, there remains some undefined quantities which we list below for the sake of completion,
\begin{align}
    \mathcal{D}_{\mathrm{n}2}^{\mu\nu\alpha} &= - \left( G_{J,\,10}^{+} \right)^{-1} \frac{2 K_{41}^{+}}{\left( \epsilon + P \right)} \left[ B^{[\mu\alpha} \left( u_\gamma \omega^{\nu]\gamma} \right) + B_{\gamma\alpha} u^{[\mu} \omega^{\nu]\gamma} - 2 \widetilde{G}_{J,\,30}^{+} u^{[\mu} \mathcal{C}^{\nu]\alpha}_{\mathrm{n}2} \right], \label{D_n2} \\
    \mathcal{C}_{\Pi}^{\mu} &= \left( \mathcal{G}_{J,\, 10}^{+} \right)^{-1} \left( \beta \mathcal{G}_{K,\, 21}^{+} - \frac{2}{3} \beta K^{+}_{42} + \mathcal{G}_{K,\, 20}^{+} \beta_\theta - \mathcal{G}_{K,\, 10}^{-} \xi_\theta \right) \left( u_\alpha \omega^{\mu\alpha} \right), \label{C_Pi} \\
    \mathcal{C}_{\mathrm{n}1}^{\mu\alpha} &= - \left( \mathcal{G}_{J,\, 10}^{+} \right)^{-1} \left\{\frac{n_{\rm f}\, K_{41}^{+} }{\left( \epsilon + P \right)} -  K_{31}^{-} \right\} \Delta^{\mu}_{\nu} \omega^{\nu\alpha}, \label{C_n1} \\
    \mathcal{C}_{\mathrm{n}2}^{\mu\alpha} &= \left( \mathcal{G}_{J,\, 10}^{+} \right)^{-1} \frac{K_{41}^{+}}{\left( \epsilon + P \right)} \Delta^{\mu\nu} \omega_{\nu\gamma} B^{\gamma\alpha}, \label{C_n2} \\
    \mathcal{C}_{\pi}^{\alpha} &= - 2 \left( \mathcal{G}_{J,\, 10}^{+} \right)^{-1} \beta K^{+}_{42} \left( u_\nu \omega^{\nu\alpha} \right), \label{C_pi} \\
    \mathcal{C}_{\Sigma}^{\mu\nu} &= \left( \mathcal{G}_{J,\, 10}^{+} \right)^{-1} J_{31}^{+} \Delta^{\mu\nu}, \label{C_Sigma} \\
    \xi_\theta &= \frac{J_{30}^{+} n_{\rm f} - J_{20}^{-} \left( \epsilon \!+\! P \right) - J_{20}^{-} B_{\alpha\beta} M^{\alpha\beta}_{\rm eq}/3}{J_{20}^{-} J_{20}^{-} - J_{10}^{+} J_{30}^{+}}, \label{xi_theta} \\
    \beta_\theta &= \frac{J_{20}^{-} n_{\rm f} - J_{10}^{+} \left( \epsilon \!+\! P \right) - J_{10}^{+} B_{\alpha\beta} M^{\alpha\beta}_{\rm eq}/3}{J_{20}^{-} J_{20}^{-} - J_{10}^{+} J_{30}^{+}}. \label{beta_theta}
\end{align}

\medskip

The transport coefficients appearing in Eq.~(35) of the main text are given by,
\begin{align}
    \beta_{n\Pi}^{\langle\mu\rangle} &= - \beta \Delta^\mu_\nu A^{\nu}_{~\,\varphi} \!\left[ \frac{\epsilon_{\psi\gamma\rho\phi} M^{\psi\gamma}_{\rm eq} B^\rho}{3 \left(\epsilon + P\right)} \left( \mathfrak{q} \tau_{\rm eq} J^{+}_{11} B^{\varphi\phi} + J^{-}_{21} \Delta^{\varphi\phi} \right) + \frac{2 \chi \mathfrak{s}^2 J_{31}^{-}}{9 m} B^{\varphi\gamma} \left( u^\psi \omega_{\gamma\psi} \right) \right] \label{beta_n,Pi} \\
    \beta_{na}^{\langle\mu\rangle\alpha} &= \frac{2 \beta\, \chi \mathfrak{s}^2\, K^{-}_{21} }{3 m} \Delta^\mu_\nu A^{\nu}_{~\,\varphi}\, \epsilon^{\rho\gamma\alpha\phi} B_\phi \left( \Delta^{\varphi\vartheta} u_{\rho} + \Delta^{\varphi}_{\rho} u^\vartheta \right) \omega_{\gamma\vartheta} \label{beta_n,a} \\
    \beta_{nn}^{\langle\mu\rangle\alpha} &= 2 \Delta^\mu_\nu A^{\nu}_{~\,\varphi} \left[ \left( \frac{n_{\mathrm{f}} J^{-}_{21} }{\left(\epsilon + P\right)} \!-\! J^{+}_{11} \right) \Delta^{\alpha\varphi} + \mathfrak{q}\, \tau_{\rm eq} \left( \frac{n_{\mathrm{f}} J^{+}_{11}}{\left(\epsilon + P\right)} - J^{-}_{01} \right) B^{\alpha\varphi} \right] \label{beta_n,n} \\
    \beta_{nF}^{\langle\mu\rangle\alpha\beta} &= \frac{\beta}{\left( \epsilon + P \right)} \Delta^{\mu}_{\nu} A^{\nu}_{~\,\varphi}\, \epsilon^{\phi\gamma\rho\beta} M_{{\rm eq},\,\phi\gamma} u_\rho \left( J^{-}_{21} \Delta^{\varphi\alpha} + \mathfrak{q}\, \tau_{\mathrm{eq}} J^{+}_{11} B^{\varphi\alpha} \right) \label{beta_n,F} \\
    \beta_{n\pi}^{\langle\mu\rangle\alpha\beta} &= \beta \Delta^{\mu}_{\nu} A^{\nu}_{\varphi}\, \epsilon^{\phi\gamma\rho\beta} \bigg[ \frac{2 \chi \mathfrak{s}^2}{3 m} K_{21}^{-} g^{\varphi}_{\phi} \left( u^\psi \omega_{\gamma\psi} \right) u_\rho B^\alpha - \frac{M_{{\rm eq},\, \phi\gamma} B_\rho}{\left(\epsilon + P \right)} \left( \mathfrak{q}\, \tau_{\mathrm{eq}} B^{\varphi\alpha} J^{+}_{11} + \Delta^{\varphi\alpha} J^{-}_{21} \right) \bigg] \label{beta_n,pi} \\
    \beta_{n\Omega}^{\langle\mu\rangle\alpha\beta} &= \beta \Delta^{\mu}_{\nu} A^{\nu}_{\varphi}\, \epsilon^{\phi\gamma\rho\alpha} \bigg[ \frac{2 \chi \mathfrak{s}^2}{3 m} K_{21}^{-} g^{\varphi}_{\phi} \left(u^\psi \omega_{\gamma\psi} \right) u_\rho B^\beta + \frac{M_{{\rm eq},\,\phi\gamma} B_\rho}{\left(\epsilon + P \right)} \left( \mathfrak{q}\, \tau_{\mathrm{eq}} B^{\varphi\beta} J^{+}_{11} + \Delta^{\varphi\beta} J^{-}_{21} \right)\bigg] \label{beta_n,Omega} \\
    \beta_{n\Sigma}^{\langle\mu\rangle\alpha\beta\gamma} &= 0 \label{beta_n,Sigma} 
\end{align}
where, $B^{\mu\nu}\equiv\epsilon^{\mu\nu\alpha\beta}u_\alpha B_\beta$ and $A^{\alpha\beta}$ is a rank-2 tensor that can be expressed as,
\begin{align}
    A^{\alpha\beta} = g^{\alpha\beta} + \frac{2 \beta\, \tau_{\rm R} J^{-}_{21}}{\left(\epsilon + P\right)} B^{\alpha\beta} + \frac{2 \beta\, \tau_{\rm R} J^{-}_{21}}{\left(\epsilon + P\right)} \left( \frac{2 \beta\, \tau_{\rm R} J^{-}_{21}}{\left(\epsilon + P\right)} + \frac{\mathfrak{q}\, \tau_{\rm R} J^{+}_{11}}{J^{-}_{21}} \right) B_{\gamma}^{~\,\alpha} B^{\beta\gamma}, \label{A^ab}
\end{align}
when terms up to quadratic in magnetic field is considered.

\medskip


The transport coefficients appearing in Eq.~(36) of the main text are given by,
\begin{align}
    \beta_{\Pi\Pi} &= \frac{2 \beta\, \chi \mathfrak{s}^2}{27 m} B_{\varphi\gamma} \!\left( 3 m^2 K^{+}_{11} \omega^{\varphi\gamma} \!+\! 10 K^{+}_{32} \Delta^{\vartheta[\gamma} \omega^{\gamma]}_{~~\vartheta} \!\right) \nonumber\\
    &~~~~- \frac{2}{D_{20}} \!\left\{\!\! \left(\! J_{20}^{+} J^{+}_{31} - J_{30}^{-} J^{-}_{21} \right)\! n_{\mathrm{f}} -\! \left(\! J_{10}^{+} J^{+}_{31} - J_{20}^{-} J^{-}_{21} \right)\! \left(\!\! \left( \epsilon + P \right) \!+\! \frac{\left(B:M_{\rm eq}\right)}{3} \!\right) \!+\! 5 \beta J^{+}_{32} \!\right\} \label{beta_Pi,Pi} \\
    \beta_{\Pi a}^{\alpha} &= \frac{\epsilon^{\alpha\phi\varphi\gamma} B_\phi}{3} \!\left[\! \frac{\left( J_{10}^{+} J^{+}_{31} \!-\! J_{20}^{-} J^{-}_{21} \right)}{2 D_{20}} M_{{\rm eq},\,\varphi\gamma} - \frac{\beta\, \mathfrak{s}^2\, \chi}{m} \Big\{ m^2 \omega_{\varphi\gamma} K^{+}_{11} + 2 \left( u^\vartheta u_{[\varphi} \omega_{\gamma]\vartheta} \left( K_{31}^{+} - K_{32}^{+} \right) + 2 \omega_{\gamma\varphi} K_{32}^{+} \right) \!\Big\} \!\right] \label{beta_Pi,a} \\
    \beta_{\Pi n}^{\alpha} &= 0 \label{beta_Pi,n} \\
    \beta_{\Pi F}^{\alpha\beta} &= 0 \label{beta_Pi,F} 
\end{align}

\begin{align}
    \beta_{\Pi\pi}^{\alpha\beta} &= - \frac{\epsilon^{\vartheta\gamma\rho\alpha} B^\beta u_\rho}{3} \left[ \frac{3 \left( J_{10}^{+} J^{+}_{31} \!-\! J_{20}^{-} J^{-}_{21} \right)}{D_{20}} M_{{\rm eq},\,\vartheta\gamma} + \frac{\beta \chi \mathfrak{s}^2}{3 m} \left(3 m^2 K^{+}_{11} - 10 K_{32}^{+} \right) \omega_{\vartheta\gamma} \right] \label{beta_Pi,pi} \\
    \beta_{\Pi\Omega}^{\alpha\beta} &= \frac{\epsilon^{\vartheta\gamma\rho\alpha} u_\rho B^\beta}{3} \left[ \frac{3 \left( J_{10}^{+} J^{+}_{31} \!-\! J_{20}^{-} J^{-}_{21} \right)}{D_{20}} M_{{\rm eq},\,\vartheta\gamma} - \frac{\beta\, \mathfrak{s}^2 \chi}{3 m} \left( 3 m^2 K^{+}_{11} + 10 K^{+}_{32} \right) \omega_{\vartheta\gamma} \right] \label{beta_Pi,Omega} \\
    \beta_{\Pi\Sigma}^{\alpha\beta\gamma} &= 0 \label{beta_Pi,Sigma} 
\end{align}

\medskip


The transport coefficients appearing in Eq.~(37) of the main text are given by,
\begin{align}
    \beta_{\pi\Pi}^{\langle\mu\nu\rangle} &= - \frac{8\, \beta\, \chi\, \mathfrak{s}^2}{9\, m} K^{+}_{32}\, B^{\gamma\langle\mu} \omega^{\nu\rangle}_{~~\gamma} \label{beta_pi,Pi} \\
    \beta_{\pi a}^{\langle\mu\nu\rangle\alpha} &= \frac{2 \beta\, \mathfrak{s}^2\, \chi}{3 m} K^{+}_{32} B_\phi \omega_{\gamma}^{~\langle\mu} \epsilon^{\nu\rangle\gamma\alpha\phi} \label{beta_pi,a} \\
    \beta_{\pi n}^{\langle\mu\nu\rangle\alpha} &= 0 \label{beta_pi,n} \\
    \beta_{\pi F}^{\langle\mu\nu\rangle\alpha\beta} &= 0 \label{beta_pi,F} \\
    \beta_{\pi\pi}^{\langle\mu\nu\rangle\alpha\beta} &= 4\, \beta\, J_{32}^{+}\, g^{\mu\alpha} g^{\nu\beta} + \frac{2 \beta \chi \mathfrak{s}^2}{3 m} K_{32}^{+} \omega_{\gamma}^{~\,\langle\mu} \epsilon^{\nu\rangle\gamma\rho\beta} u_\rho B^\alpha - 8 \mathfrak{q}\, \beta \tau_{\mathrm{eq}} J^{-}_{22} B^{\beta\langle\mu} g^{\nu\rangle\alpha} \label{beta_pi,pi} \\
    \beta_{\pi\Omega}^{\langle\mu\nu\rangle\alpha\beta} &= - \frac{4 \beta\, \mathfrak{s}^2 \chi}{3 m} K^{+}_{32}\, \omega_{\gamma}^{~\,\langle\mu} \epsilon^{\nu\rangle\gamma\rho\beta} u_\rho B^\alpha \label{beta_pi,Omega} \\
    \beta_{\pi\Sigma}^{\langle\mu\nu\rangle\alpha\beta\gamma} &= 0 \label{beta_pi,Sigma} 
\end{align}

\medskip


The transport coefficients appearing in Eq.~(38) of the main text are given by,
\begin{align}
    B_{\Pi}^{\lambda,[\mu\nu]} &= \frac{1}{2 D_{20}} \!\left\{\! J_{20}^{+} \left(m^2 \omega^{\mu\nu} u_\vartheta K^{\lambda\vartheta}_{(0)+} + 2 K^{\lambda\vartheta\psi[\mu}_{(0)+} \omega^{\nu]}_{~\,\psi} \right) - J_{30}^{-} \left(m^2 \omega^{\mu\nu} K^{\lambda}_{(0)-} + 2 K^{\lambda\psi[\mu}_{(0)+} \omega^{\nu]}_{~\,\psi} \right) \right\} n_{\mathrm{f}} \nonumber\\
    &~~-\! \frac{1}{2 D_{20}} \!\left\{\! J_{10}^{+} \left(m^2 \omega^{\mu\nu} u_\vartheta K^{\lambda\vartheta}_{(0)+} + 2 K^{\lambda\vartheta\psi[\mu}_{(0)+} \omega^{\nu]}_{~\,\psi} \right) - J_{20}^{-} \left(m^2 \omega^{\mu\nu} K^{\lambda}_{(0)-} + 2 K^{\lambda\psi[\mu}_{(0)-}  \omega^{\nu]}_{~\,\psi} \right) \right\} \nonumber\\
    &~~\times \left\{\! \left( \epsilon + P \right) + \frac{\left(B:M_{\rm eq}\right)}{3} \right\} + \frac{\beta}{6} \Delta_{\alpha\phi} \left(m^2 \omega^{\mu\nu} K^{\lambda\alpha\phi}_{(1)+} + 2 K^{\lambda\alpha\phi\psi[\mu}_{(1)+} \omega^{\nu]}_{~\,\psi} \right) \nonumber\\
    &~~- \frac{1}{2} u_\vartheta \left(m^2 \mathcal{D}_{\Pi}^{[\mu\nu]} J^{\lambda\vartheta}_{(1)+} + 2 J^{\lambda\vartheta[\mu}_{(1)+,\,\psi} \mathcal{D}_{\Pi}^{\nu]\psi} \right) - \frac{\beta\, m\, \chi}{3} \left( m^2 \epsilon^{\mu\nu\rho\phi} J^{\lambda}_{(1)+} + 2 J^{\lambda[\mu}_{(1)+,\,\psi} \epsilon^{\nu]\psi\rho\phi} \right) u_\rho B_\phi \nonumber\\
    &~~- \frac{\beta \mathfrak{q}\, \tau_{\mathrm{R}}}{3} B_{\phi\beta} \left(m^2 \omega^{\mu\nu} K^{\lambda\beta\phi}_{(2)-} + 2 K^{\lambda\beta\phi\psi[\mu}_{(2)-} \omega^{\nu]}_{~\,\psi} \right) + A^{\lambda[\mu\nu]}_{~~~~\,\phi}\, \beta_{n\Pi}^{\langle\phi\rangle} \label{B_Pi} \\
    B_{a}^{\lambda,[\mu\nu]\alpha} &= \frac{\beta\, m\, \chi}{2} B_\phi \left( m^2 J^{\lambda}_{(1)+} \epsilon^{\mu\nu\alpha\phi} + 2 J^{\lambda\gamma[\mu}_{(1)+} \epsilon^{\nu]\alpha\phi}_{~~~~\,\gamma} \right) \Dot{u}_\alpha + A^{\lambda[\mu\nu]}_{~~~~\,\phi}\, \beta_{na}^{\langle\phi\rangle\alpha} \label{B_a} \\
    B_{n}^{\lambda,[\mu\nu]\alpha} &= \frac{1}{2} \bigg[\! \frac{n_{\mathrm{f}} }{\left(\epsilon + P\right)} \left( m^2 \omega^{\mu\nu} K^{\lambda\alpha}_{(0)+} + 2 K^{\lambda\alpha\psi[\mu}_{(0)+} \omega^{\nu]}_{~\,\psi} \right) \!-\! \left( m^2 \omega^{\mu\nu} K^{\lambda\alpha}_{(1)-} + 2 K^{\lambda\alpha\psi[\mu}_{(1)-} \omega^{\nu]}_{~\,\psi} \right) \nonumber\\
    &~~- \left( m^2 \mathcal{D}_{\mathrm{n}}^{[\mu\nu]\alpha} J^{\lambda}_{(0)+} + 2 J^{\lambda[\mu}_{(0)+,\,\psi} \mathcal{D}_{\mathrm{n}}^{\nu]\psi\alpha} \right) \bigg] - \frac{\mathfrak{q}\, \tau_{\mathrm{eq}}}{2} B^{\alpha}_{~\,\rho} \bigg[\! \frac{n_{\mathrm{f}} }{\left(\epsilon + P\right)} \left( m^2 \omega^{\mu\nu} K^{\lambda\rho}_{(1)-} + 2 K^{\lambda\rho\psi[\mu}_{(1)-} \omega^{\nu]}_{~\,\psi} \right) \nonumber\\
    &~~- \left( m^2 \omega^{\mu\nu} K^{\lambda\rho}_{(2)+} + 2 K^{\lambda\rho\psi[\mu}_{(2)+} \omega^{\nu]}_{~\,\psi} \right) \bigg] + A^{\lambda[\mu\nu]}_{~~~~\,\phi}\, \beta_{nn}^{\langle\phi\rangle\alpha} \label{B_n} \\
    B_{F}^{\lambda,[\mu\nu]\alpha\beta} &= 0 \label{B_F} \\
    B_{\pi}^{\lambda,[\mu\nu]\alpha\beta} &=\! \frac{\beta}{2} \!\left( m^2 \omega^{\mu\nu} K^{\lambda\alpha\beta}_{(1)+} + 2 K^{\lambda\alpha\beta\psi[\mu}_{(1)+} \omega^{\nu]}_{~\,\psi} \right) \!-\! \frac{1}{2} \left( m^2 \mathcal{D}_{\pi}^{[\mu\nu]\alpha\beta} J^{\lambda}_{(0)+} + 2 J^{\lambda[\mu}_{(0)+,\,\psi} \mathcal{D}_{\pi}^{\nu]\psi\alpha\beta} \right) \nonumber\\
    &~~- \beta \mathfrak{q}\, \tau_{\mathrm{eq}} B^{\alpha}_{~\,\varphi} \left( m^2 \omega^{\mu\nu} K^{\lambda\varphi\beta}_{(2)-} + 2 K^{\lambda\varphi\beta\psi[\mu}_{(2)-} \omega^{\nu]}_{~\,\psi} \right) + A^{\lambda[\mu\nu]}_{~~~~\,\phi}\, \beta_{n\pi}^{\langle\phi\rangle\alpha\beta} \label{B_pi} \\
    B_{\Omega}^{\lambda,[\mu\nu]\alpha\beta} &= - \frac{m \beta \chi}{2} u_\phi B^\alpha \!\left( m^2 \epsilon^{\mu\nu\phi\beta} J^{\lambda}_{(1)+} \!+\! 2 J^{\lambda\psi[\mu}_{(1)+} \epsilon^{\nu]\phi\beta}_{~~~~\,\psi} \right) + A^{\lambda[\mu\nu]}_{~~~~\,\phi}\, \beta_{n\Omega}^{\langle\phi\rangle\alpha\beta} \label{B_Omega} 
\end{align} 
    
\begin{align}
    B_{\Sigma}^{\lambda,[\mu\nu]\alpha\beta\gamma} &= - \frac{1}{2} \bigg[ \left( m^2 J^{\lambda}_{(0)+} \mathcal{D}_{\Sigma}^{[\mu\nu]\alpha\beta\gamma} + 2 J^{\lambda[\mu}_{(0)+,\,\phi} \mathcal{D}_{\Sigma}^{\nu]\phi\alpha\beta\gamma} \right) + \left( m^2 J^{\lambda\alpha}_{(1)+} g^{\beta[\mu} g^{\nu]\gamma} + 2 J^{\lambda\alpha\gamma[\mu}_{(1)+} g^{\nu]\beta} \right) \nonumber\\
    &~~- \mathfrak{q}\, \tau_{\mathrm{eq}} B^{\alpha}_{~\,\phi} \left( m^2 J^{\lambda\phi}_{(2)-} g^{\beta[\mu} g^{\nu]\gamma} + 2 J^{\lambda\phi\gamma[\mu}_{(2)-} g^{\nu]\beta} \right) \bigg] + A^{\lambda[\mu\nu]}_{~~~~\,\phi}\, \beta_{n\Omega}^{\langle\phi\rangle\alpha\beta\gamma} \label{B_Sigma}
\end{align}
where, $A^{\lambda\mu\nu\phi}$ is a rank-4 tensor that can be expressed as,
\begin{align}
    A^{\lambda\mu\nu\phi} &= \frac{1}{2} \bigg[\! \frac{\beta B_{\alpha}^{~\,\phi}}{\left(\epsilon + P\right)} \!\left( m^2 \omega^{\mu\nu} K^{\lambda\alpha}_{(0)+} \!+\! 2 K^{\lambda\alpha\psi[\mu}_{(0)+} \omega^{\nu]}_{~\,\psi} \right) \!-\! \left( m^2 \mathcal{D}_{\mathrm{n2}}^{\mu\nu\phi} J^{\lambda}_{(0)+} \!+\! 2 J^{\lambda[\mu}_{(0)+,\,\psi} \mathcal{D}_{\mathrm{n2}}^{\nu]\psi\phi} \right) \nonumber\\
    &~~-\! \frac{\beta \mathfrak{q}\, \tau_{\mathrm{eq}}}{\left(\epsilon + P\right)} B_{\alpha\beta} B^{\alpha\phi} \!\left( m^2 \omega^{\mu\nu} K^{\lambda\beta}_{(1)-} \!+\! 2 K^{\lambda\beta\psi[\mu}_{(1)-} \omega^{\nu]}_{~\,\psi} \right) \!\!\bigg] \label{A^lmnp}
\end{align}

\vspace{0.2cm}
\begin{center}
\underline{\textbf{Entropy production and dissipative gradients}}
\end{center}

Using Boltzmann H-theorem, the entropy four-current can be written in terms of the distribution function and simplified as
\begin{align}
    {\cal H}^\mu &= - \int \mathrm{dP} \mathrm{dS} p^\mu \left[ \left( f \ln f + \Tilde{f} \ln\Tilde{f} \right) + \left( \Bar{f} \ln \Bar{f} + \Tilde{\Bar{f}} \ln\Tilde{\Bar{f}} \right) \right] \label{S^mu1} \\
    &= \int \mathrm{dP} \mathrm{dS} p^\mu \left[ \beta \left( u \cdot p \right) \left( f + \Bar{f} \right) - \xi \left( f - \Bar{f} \right) - \frac{1}{2} \left(s:\omega\right) \left( f + \Bar{f} \right) \right]  - \int \mathrm{dP} \mathrm{dS} p^\mu \left( \ln\Tilde{f}_0 + \ln\Tilde{\Bar{f}}_0 \right)\label{S^mu2} \\
    &= \beta_\nu T^{\mu\nu}_{\rm f} - \xi N^\mu - \frac{1}{2} \omega_{\alpha\beta} S^{\mu,\alpha\beta} + P \beta^\mu \label{S^mu3}
\end{align}
Taking divergence of the entropy four-current, we get
\begin{equation}\label{SMHD entropy production}
\partial_\mu {\cal H}^\mu = \beta \sigma_{\mu\nu} \pi^{\mu\nu} - \beta \theta \Pi - \left( \nabla_\mu \xi \right) n^\mu - \frac{1}{2} \left( \nabla_\mu \omega_{\alpha\beta} \right) \delta S^{\mu,\alpha\beta}
\end{equation}
To arrive at the above relation, we have used the matching conditions defined in the main text and combined results of Refs.~[41,80]
\begin{align}
 P \beta^\mu &=- \int \mathrm{dP} \mathrm{dS} p^\mu \left( \ln\Tilde{f}_0 + \ln\Tilde{\Bar{f}}_0 \right)\\
 \partial_\mu\left( P \beta^\mu \right) &= -\left( \partial_\mu \beta_\nu \right) T^{\mu\nu}_{\rm f,eq} + \left( \partial_\mu \xi \right) N^\mu_{\rm eq} + \frac{1}{2} \left( \partial_\mu \omega_{\alpha\beta} \right) S^{\mu,\alpha\beta}_{\rm eq} - \frac{1}{2} \beta_\nu \left( \partial^\nu F^{\alpha\beta} \right) M_{\alpha\beta} 
\end{align}
Finally, from Eq.~\eqref{SMHD entropy production}, we conclude that positive definiteness of entropy production leads to the following form of the dissipative quantities
\begin{align}
    \Pi &= - \zeta \theta \label{Disp_curr1}\\
    n^{\mu} &= \kappa^{\mu\alpha} \left( \nabla_{\alpha} \xi \right) \label{Disp_curr2}\\
    \pi^{\mu\nu} &= \eta^{\mu\nu\alpha\beta} \sigma_{\alpha\beta} \label{Disp_curr3}\\
    \delta S^{\mu,\alpha\beta} &= \Sigma^{\mu\alpha\beta\lambda\gamma\rho} \left(\nabla_\lambda \omega_{\gamma\rho} \right) \label{Disp_curr4}
\end{align}

\end{document}